\pdfoutput=1

\documentclass[twocolumn,showpacs]{revtex4}
\usepackage{amssymb,amsmath,graphicx}
\usepackage{color}
\begin{document}

\title{
  Magnetoconductance of the Corbino disk in graphene: 
  Chiral tunneling and quantum interference in the bilayer case 
}

\author{Grzegorz Rut}
\affiliation{Marian Smoluchowski Institute of Physics, 
Jagiellonian University, Reymonta 4, PL--30059 Krak\'{o}w, Poland}
\author{Adam Rycerz}
\affiliation{Marian Smoluchowski Institute of Physics, 
Jagiellonian University, Reymonta 4, PL--30059 Krak\'{o}w, Poland}

\begin{abstract}
Quantum transport through an impurity-free Corbino disk in bilayer graphene is investigated analytically, by the mode-matching method for effective Dirac equation, in the presence of uniform magnetic fields. Similarly as in the monolayer case (see Refs.\ \cite{Ryc10,Kat10}), conductance at the Dirac point shows oscillations with the flux piercing the disk area $\Phi_D$ characterized by the period $\Phi_0=2\,(h/e)\ln(R_\mathrm{o}/R_\mathrm{i})$, where $R_\mathrm{o}$ ($R_\mathrm{i}$) is the outer (inner) disk radius. The oscillations magnitude depends either on the radii ratio or on the physical disk size, with the condition for maximal oscillations reading $R_{\rm o}/R_{\rm i}\simeq\left[\,R_{\rm i}t_{\perp}/(2\hbar{}v_{F})\,\right]^{4/p}$ (for $R_{\rm o}/R_{\rm i}\gg{}1$), where $t_\perp$ is the interlayer hopping integral, $v_F$ is the Fermi velocity in graphene, and $p$ is an {\em even} integer. {\em Odd}-integer values of $p$ correspond to vanishing oscillations for the normal Corbino setup, or to oscillations frequency doubling for the Andreev-Corbino setup. At higher Landau levels (LLs) magnetoconductance behaves almost identically in the monolayer and bilayer cases. A brief comparison with the Corbino disk in 2DEG is also provided in order to illustrate the role of chiral tunneling in graphene. 
\end{abstract}

\date{October 2, 2014}
\pacs{  72.80.Vp, 73.43.Qt, 73.63.-b, 75.47.-m  }
\maketitle

\section{Introduction}
\begin{figure}
\centerline{\includegraphics[width=0.9\linewidth]{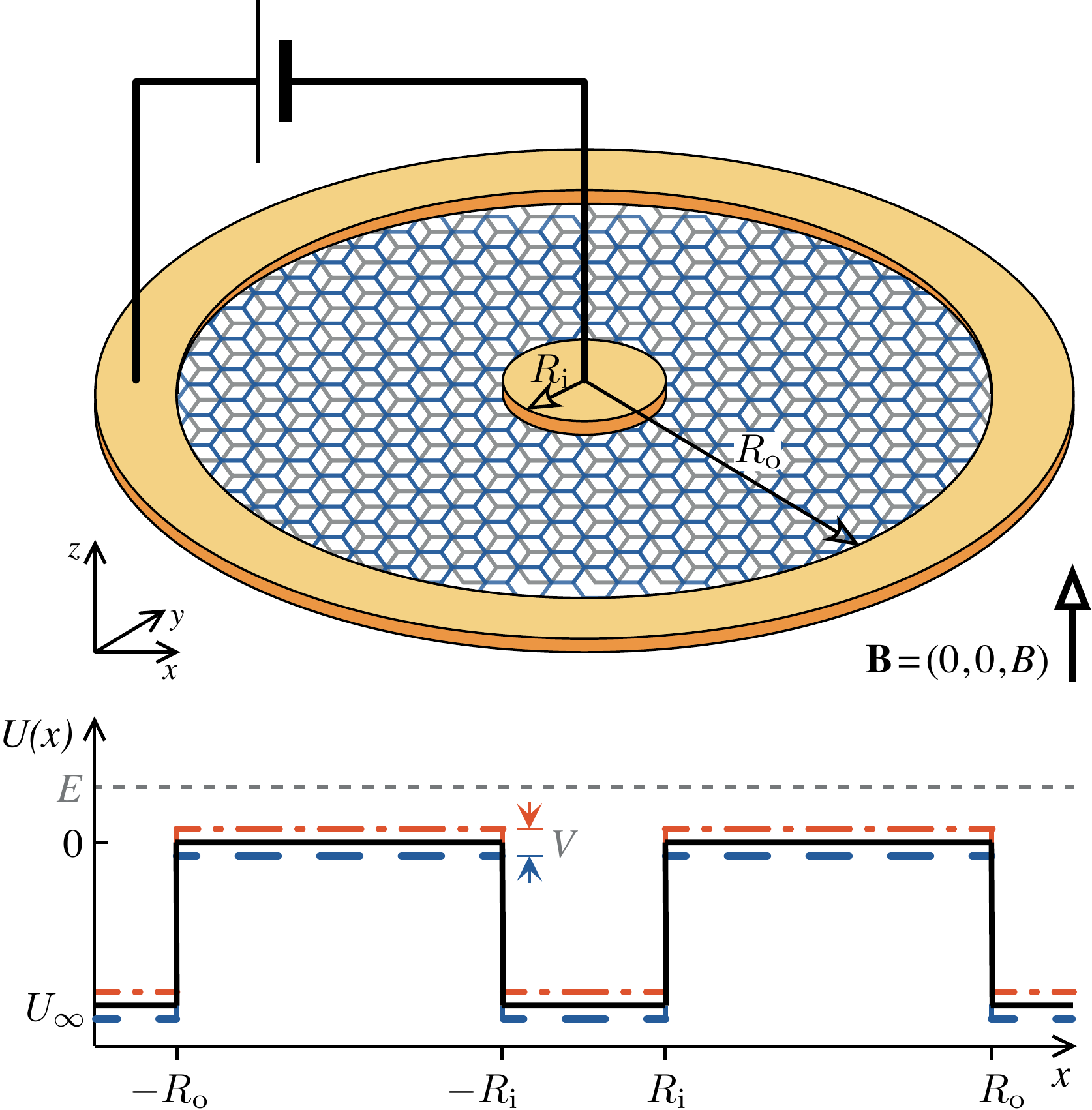}}
\caption{\label{bicorbino}
  The Corbino disk in $AB$ stacked bilayer graphene. Top: Device schematics. The current is passed through the disk-shaped area with the inner radius $R_\mathrm{i}$ and the outer radius $R_\mathrm{o}$ in a~perpendicular magnetic field $\mathbf{B}=(0,0,B)$. Bottom: The electrostatic potential cross section along the $x$-axis $U(x)$ following from Eq.\ (\ref{eq:potential}) [black solid line]. The leads (yellow areas) are modeled as infinitely-doped graphene regions ($|U_\infty|\rightarrow\infty$). The additional top- and bottom-gate electrodes (not shown) are used to tune the Fermi energy ($E$) [grey dotted line] and to induce the electrostatic bias between the layers ($V$), leading to the local potential energies $U(x)+V/2$ [red dash-dot line] and $U(x)-V/2$ [blue dashed line]. 
}
\end{figure}

The potential of bilayer graphene (BLG) for carbon-based electronics rests on the possibility to control its transport properties by external electromagnetic fields employing the mechanisms that have no analogues in monolayer graphene (MLG) or in semiconducting heterostructures containing two-dimensional electron gas (2DEG) \cite{Mac13}. BLG with $AB$ stacking order can be converted from a~semimetal to a~narrow gap semiconductor by applying a~perpendicular electrostatic field \cite{Mac06,Oht06,Oos07,Cas07,Per07a}. This is possible, because $(i)$ the interlayer hoppings break the sublattice symmetry in a~single layer, leading to the formation of two parabolic chiral bands touching themselves at the so-called Dirac points \cite{Mac06,footriwa}, and $(ii)$ the perpendicular electric field further breaks the inversion symmetry, opening a~gap between conduction and valence bands. Several experiments on dual-gated devices in ultraclean BLG have pursued the possibility of exploiting such a~field-tunable energy gap \cite{Fel09,Wei10,Yan10,May11,Rtt11,Bao12,Vel12}. Yan and Fuhrer \cite{Yan10} used the Corbino geometry, proposed over a~century ago to measure the magnetoresistance without generating the Hall voltage \cite{Gal91}. In such a~geometry (see Fig.\ \ref{bicorbino}) the current is passed through a~disk-shaped sample surrounded from both exterior and interior sides with metallic leads, suppressing the influence of boundary modes \cite{Zha13} on various dynamical properties of nanosystems in both BLG and MLG \cite{Fau10,Liu11,Zha12,Pet14}. 

From a~more fundamental point of view, several relativistic quantum effects, observed for MLG and resulting from the chiral nature of effective quasiparticles, are predicted to manifest themselves in BLG in slightly modified versions, mainly due to the presence of new characteristic length scale for low-energy excitations \cite{Sny07}
\begin{equation}
  \label{lperp}
  l_\perp=\hbar{}v_F/t_\perp\simeq{}11\,d_0,
\end{equation}
where $v_F\simeq{}10^6\,$m/s is the energy-independent Fermi velocity in MLG, $t_\perp\simeq{}0.4\,$eV is the nearest-neighbor interlayer hopping integral, and $d_0=0.142\,$nm is a~C-C bond length. For instance, the universal ballistic conductivity of MLG $\sigma_0=(4/\pi)\,e^2/h$, characterizing the so-called {\em pseudodiffusive} transport regime \cite{Kat06a,Two06,Mia07,Dan08,Ryc09}, is replaced by the length-dependent value $\sigma(L)$ varying from $\sigma_0$ to $3\sigma_0$ per layer \cite{Mog09,Nil08,Rut14a}, with the upper limit approached for the system size $L\rightarrow\infty$. In the quantum-Hall regime, the zero-energy Landau level ($0$LL) shows the eightfold degeneracy for BLG (instead of the fourfold degeneracy for MLG) which can be lifted by manipulating the external electromagnetic fields, partly due to a~role of electron-electron interactions \cite{Wei10,Rtt11,Zha12}. Also, the quantum-interference in graphene Aharonov-Bohm rings \cite{Sch12} may result in different oscillation patterns appearing for MLG and BLG cases \cite{Xav10,Zar10,Rom12}.

An intriguing quantum-interference phenomenon was predicted theoretically for impurity-free Corbino disks in MLG \cite{Ryc10,Kat10,Ryc12,Kha13,Bah13,Rut14b}. In brief, periodic (approximately sinusoidal) magnetoconductance oscillations are followed, for an {\em undoped} sample, by similar oscillations of the shot-noise power \cite{Kat10} and the third charge-transfer cumulant \cite{Ryc12,Rut14b}. The effect has a~direct analog for strain-induced pseudomagnetic fields \cite{Kha13}, allowing to consider a~fully-mesoscopic counterpart to the earlier proposed valley filters in MLG \cite{Ryc07,Akh08,Gun11} or carbon nanotubes \cite{Pal11}. At higher dopings, the oscillations reappear provided the magnetic field is adjusted to the positions of $n$-th Landau level ($n$LL) in the field-doping parameter plane \cite{Ryc10}. Also very recently, LL splittings due to a~possible substrate-induced spin-orbit interaction in Corbino devices were discussed as an alternative mechanism \cite{Vil14} for graphene-based spintronics \cite{Wim08}.

Most remarkably, the disk conductance averaged over a~single period restores the pseudodiffusive value \cite{Ryc09}
\begin{equation}
  \label{gdiffmono}
  G_{\it diff}^{\rm MLG}=\frac{2\pi\sigma_0}{\ln\left(R_{\rm o}/R_{\rm i}\right)}.
\end{equation}
Analogous behavior is predicted for higher charge-transfer cumulants \cite{foofinvo}, showing that the effect is another manifestation of the chiral nature of Dirac fermions in graphene. For these reasons, we have coined the~term of {\em quantum-relativistic Corbino effect} (QRCE). 

In this paper, magnetoconductance of the Corbino disk in BLG is discussed in analytical terms, starting from the four-band effective Hamiltonian \cite{Mac06} and employing the Landauer-B\"{u}ttiker formalism \cite{Naz09} for the linear-response regime. The paper is organized as follows. In Sec.\ II we present the system details and discuss the solutions of the corresponding Dirac equation for arbitrary dopings and magnetic fields. Then, in Sec.\ III, magnetotransport signatures of QRCE are demonstrated for the normal Corbino and for the Andreev-Corbino setup. Sec.\ IV provides a~quantitative comparison with the magnetoconductance spectra for the Corbino disk in 2DEG. The conclusions are given in Sec.\ V. 

\section{Mode-matching for the effective Dirac equation}
\subsection{The Hamiltonian and envelope wavefunctions}
The analysis starts from the four-band effective Hamiltonian for $K$ valley \cite{Mac06}, which is given by
\begin{equation}
  \label{eq:hamiltonian1}
  H=\left(\begin{array}{cccc}
      V/2 & \pi & t_{\bot} & 0\\
      \pi^\dagger & V/2 & 0 & 0\\
      t_{\bot} & 0 & -V/2 & \pi^\dagger\\
      0 & 0 & \pi & -V/2
    \end{array}\right) + U(r)\,\mathbb{I},
\end{equation}
where $V$ is the electrostatic bias between the layers $\pi=\pi_x+i\pi_y$ and  $\pi^\dagger=\pi_x-i\pi_y$, with $\pi_j/v_{F}=\left(-i\hbar\,\partial_j+eA_j\right)$ being a~component of the gauge-invariant momentum operator ($j=1,2$), the electron charge is $-e$, the potential energy term $U(r)\,\mathbb{I}$ depends only on $r=\sqrt{x^2+y^2}$ (with $\mathbb{I}$ the identity matrix), and the remaining symbols are same as in Eq.\ (\ref{lperp}). We choose the symmetric gauge ${\bf A}\equiv{}(A_x,A_y)=(B/2)\left(-y,x\right)$, with the uniform magnetic field $B\neq{}0$ in the disk area ($R_\mathrm{i}<r<R_\mathrm{o}$) and $B=0$ otherwise. The inner and outer contacts are modeled with heavily doped BLG areas; that is, we set the potential energy profile in Eq.\ (\ref{eq:hamiltonian1}) as follows
\begin{equation}
  \label{eq:potential}
  U(r)=\begin{cases}
    U_{\infty} & \mbox{if \ensuremath{r<R_{\rm i}\mbox{ or }r>R_{\rm o}},}\\
    0 & \mbox{if \ensuremath{R_\mathrm{i}<r<R_\mathrm{o}}},
  \end{cases}
\end{equation}
and focus on the limit of $|U_\infty|\rightarrow\infty$. In order to obtain the Hamiltonian for the other valley ($K'$), it is sufficient to substitute $V\rightarrow{}-V$ and $\pi\rightarrow{}-\pi$ in Eq.\ (\ref{eq:hamiltonian1}).

Since our model system possesses a~cylindrical symmetry, the Hamiltonian (\ref{eq:hamiltonian1}) commutes with the total angular-momentum operator \cite{Per07b}
\begin{equation}
  \label{jztotal}
  J_z=-i\hbar\partial_\varphi + \frac{\hbar}{2}
  \left(\begin{array}{cc}
      \sigma_0 & 0 \\
      0 & -\sigma_0
    \end{array}\right) + \frac{\hbar}{2}
  \left(\begin{array}{cc}
      -\sigma_z & 0 \\
      0 & \sigma_z
    \end{array}\right),
\end{equation}
where $\sigma_0$ is the $2\times{}2$ identity matrix, $\sigma_z$ is one of the Pauli matrices, and we have used the polar coordinates $(r,\varphi)$. In turn, the wavefunctions can be written as products of angular and radial parts (the so-called {\em envelope wavefunctions}), namely
\begin{equation}
  \label{eq:fi}
  \psi\left(r,\varphi\right)=e^{im\varphi}\left(\begin{array}{c}
      \phi_{1}(r)\\
      ie^{-i\varphi}\phi_{2}(r)\\
      \phi_{3}(r)\\
      ie^{i\varphi}\phi_{4}(r)
    \end{array}\right)
\end{equation}
where $m=0,\pm{}1,\pm{}2,\dots$. Notice that the angular momentum quantum number in BLG case is an integer $m$, in contrast to the half-odd integer $j$ in MLG case \cite{Rec09a}.

\subsection{The contact regions}
For the contact regions ($r<R_{\rm i}$ or $r>R_{\rm o}$), we have $B=0$ and thus the four-band Dirac equation $H\psi=E\psi$, with $H$ given by Eq.\ (\ref{eq:hamiltonian1}) and $E$ being the Fermi energy, can be written as
\begin{equation}
  \label{eq:diaceq}
  \left(\begin{array}{cccc}
      \tilde{\epsilon}+\Delta & \kappa_{+} & -l_{\perp}^{-1} & 0\\
      \kappa_{-} & \tilde{\epsilon}+\Delta & 0 & 0\\
      -l_{\perp}^{-1} & 0 & \tilde{\epsilon}-\Delta & \kappa_{-}\\
      0 & 0 & \kappa_{+} & \tilde{\epsilon}-\Delta
    \end{array}\right)\psi\left(r,\phi\right)=0
\end{equation}
where $\tilde{\epsilon}=(E-U_\infty)/\left(\hbar v_{F}\right)$, $\kappa_{\pm}=ie^{\pm i\varphi}\left(\partial_{r}\pm ir^{-1}\partial_{\varphi}\right)$, and $\Delta=-V/\left(2\hbar v_{F}\right)$. 
Substituting $\psi\left(r,\varphi\right)$ (\ref{eq:fi}) into Eq.\ (\ref{eq:diaceq}) and decoupling the equation for $\phi_{1}^{\pm}(r)$ one gets 
\begin{equation}
  \label{phi1leads}
  \left(
    \partial_{r}^{2}+\frac{1}{r}\partial_{r}-\frac{m^{2}}{r^{2}}+\eta_{\pm}
  \right)\phi_{1}^{\pm}(r)=0,
\end{equation}
where $\eta_{\pm}=\left(\Delta^{2}+\tilde{\epsilon}^{2}\right)\pm\sqrt{\tilde{\epsilon}^{2}\left(4\Delta^{2}+1/l_{\perp}^{2}\right)-\Delta^{2}/l_{\perp}^{2}}$.
Next, using the differential relations following from Eq.\ (\ref{eq:diaceq}), one can obtain the remaining components of the wavefunction $\phi^\pm(r)=\left[\phi_{1}^{\pm}(r),\phi_{2}^{\pm}(r),\phi_{3}^{\pm}(r),\phi_{4}^{\pm}(r)\right]^T$, which are given explicitly in Appendix~A. 

\subsection{The disk area}
For the disk area ($R_\mathrm{i}<r<R_\mathrm{o}$), we have $B\neq{}0$ and it is convenient to define the dimensionless variable $\rho=r/l_{B}$, with the magnetic length $l_{B}=\sqrt{\hbar/|eB|}$. In turn, Eq.\ (\ref{eq:diaceq}) is replaced by
\begin{equation}
  \label{eq:diaceq1}
  \left(\begin{array}{cccc}
      \varepsilon+\delta & \xi_{+} & -t & 0\\
      \xi_{-} & \varepsilon+\delta & 0 & 0\\
      -t & 0 & \varepsilon-\delta & \xi_{-}\\
      0 & 0 & \xi_{+} & \varepsilon-\delta
    \end{array}\right)\psi\left(\rho,\phi\right)=0,
\end{equation}
where $t=l_{B}/l_{\perp}$, $\varepsilon=El_{B}/(\hbar v_{F})$, $\delta=-Vl_{B}/(2\hbar v_{F})$,
$\xi_{\pm}=i\,\mbox{exp}\left(\pm i\varphi\right)\left(\partial_{\rho}\pm i\rho^{-1}\,\partial_{\varphi}\mp\rho/2\right)$. Eliminating the angle-dependent part of the wavefunction, we obtain
\begin{equation}
  \label{phi1disk}
  \left(
    \partial_{\rho}^{2}+\frac{1}{\rho}\partial_{\rho}
    -\frac{\rho^{2}}{4}-\frac{m^{2}}{\rho^{2}}-m\!-\!1+\gamma_{\pm}
  \right)\phi_{1}^{\pm}\left(\rho\right) = 0,
\end{equation}
where $\gamma_{\pm}=\left(\delta^{2}+\varepsilon^{2}\right)\pm\sqrt{\varepsilon^{2}\left(4\delta^{2}+t{}^{2}\right)-\delta^{2}t^{2}}$.
The complete solution of Eq.\ (\ref{eq:diaceq1}) is presented in Appendix~A. It can be shown that the normalization condition for the wavefunction leads to the energies of Landau levels \cite{Per07a}
\begin{equation}
  \label{eq:llBLG}
  \gamma_{\pm}=n+\frac{|m|+m+1}{2},
\end{equation}
where $n=0,1,2$...

\subsection{Reflection and transmission coefficients}
Next, we consider the scattering problem for the radial wave functions, assuming that the initial wave is incoming from the inner lead. The solutions of Eq.\ (\ref{eq:diaceq}) for the inner and outer lead can be presented as follows
\begin{align}
  \phi_{\rm i}^\pm(r) &= 
  \phi_{\sf in}^\pm(r) + r_p^\pm\phi_{\sf out}^+(r) + r_n^\pm\phi_{\sf out}^-(r),
  \label{phiilin} \\
  \phi_{\rm o}^\pm(r) &= t_p^\pm\phi_{\sf in}^+(r) + t_n^\pm\phi_{\sf in}^-(r), 
  \label{phiolin} 
\end{align}
where $\phi_{\sf in}^\pm(r)$ and $\phi_{\sf out}^\pm(r)$ denotes the wavefunctions propagating from $r=0$ and $r=\infty$ (respectively) and carrying the unit current. In analogy, a~general solution of Eq.\ (\ref{eq:diaceq1}) for the disk area corresponds to the linear combination of four eigenspinors, namely
\begin{equation}
  \label{phidlin}
  \phi_{\rm d}^\pm(r) = \sum_{\mu=1}^4\alpha_\mu^\pm\phi_\mu(r),
\end{equation}
where $\{\alpha_\mu^\pm\}_{\mu=1\dots{}4}$ are arbitrary complex coefficients. Matching the wavefunctions $\phi_{\rm i}^\pm(r)$ (\ref{phiilin}) and $\phi_{\rm d}^\pm(r)$ (\ref{phidlin}) at $r=R_{\rm i}$, as well as $\phi_{\rm d}^\pm(r)$ (\ref{phidlin}) and $\phi_{\rm o}^\pm(r)$ (\ref{phiolin}) at $r=R_{\rm o}$, we obtain the  reflection and transmission coefficients corresponding to the $K$ valley and the angular momentum quantum number $m$, which can be arranged in $2\times{}2$ matrices, namely
\begin{equation}
\label{rt1block}
  \mathbf{r}_{K,m}=\left(\begin{array}{cc}
      r_{p}^{+} & r_{n}^{+}\\
      r_{p}^{-} & r_{n}^{-}
    \end{array}\right),\ \ \ \ 
  \mathbf{t}_{K,m}=\left(\begin{array}{cc}
      t_{p}^{+} & t_{n}^{+}\\
      t_{p}^{-} & t_{n}^{-}
    \end{array}\right).
\end{equation}
The remaining details of the mode matching procedure are given in Appendix~B. 

It is worth to mention here, that skew-interlayer hoppings \cite{Mac13}, neglected in the Hamiltonian $H$ (\ref{eq:hamiltonian1}), are predicted theoretically to enhance, typically by a~factor of $3$, the zero-magnetic field conductivity of large bilayer samples at the Dirac point \cite{Mac13,Mog09,Cse07}. The experimental value reported by Ref.\ \cite{May11} are close, but noticeably smaller than the theoretical prediction, what can be attributed to the several factors, including the finite system size \cite{Rut14a}. Nevertheless, it is also shown in Ref.\ \cite{Rut14a} that the conductance of finite bilayer samples (with the length $L\lesssim{}100\,$nm) becomes insensitive to skew-interlayer hoppings at high magnetic fields $B\gtrsim{}5\,$T, and thus the scattering approach constituted by the four-band Hamiltonian (\ref{eq:hamiltonian1}) is sufficient to discuss basic magnetotransport characteristics of nanoscale devices in BLG.

\section{ \label{seconqrce}
  Quantum relativistic Corbino effect (QRCE) in BLG}
In this section we present our main results concerning the magnetoconductance of the Corbino disk in BLG. In the linear-response regime, the conductance is given by the Landauer-B\"{u}ttiker formula \cite{Lan70} 
\begin{equation}
  \label{gland}
  G=2_{s}2_{v}g_{0}\mbox{Tr}\,\boldsymbol{T},
\end{equation}
where $g_{0}=e^{2}/h$ is the conductance quantum, $2_{s(v)}$ is the spin
(valley) degeneracy, $\boldsymbol{T}=\boldsymbol{t^{\dagger}t}$ and $\boldsymbol{t}$ is a~block-diagonal matrix with each block given by the second equality in Eq.\ (\ref{rt1block}). The Zeeman splitting is neglected for clarity. At first step, we have also assumed the unbiased sample case ($V=0$), for which the twofold valley degeneracy occurs. The $V\neq{}0$ case is discussed separately later in this section.

\begin{figure}
\centerline{\includegraphics[width=0.8\linewidth]{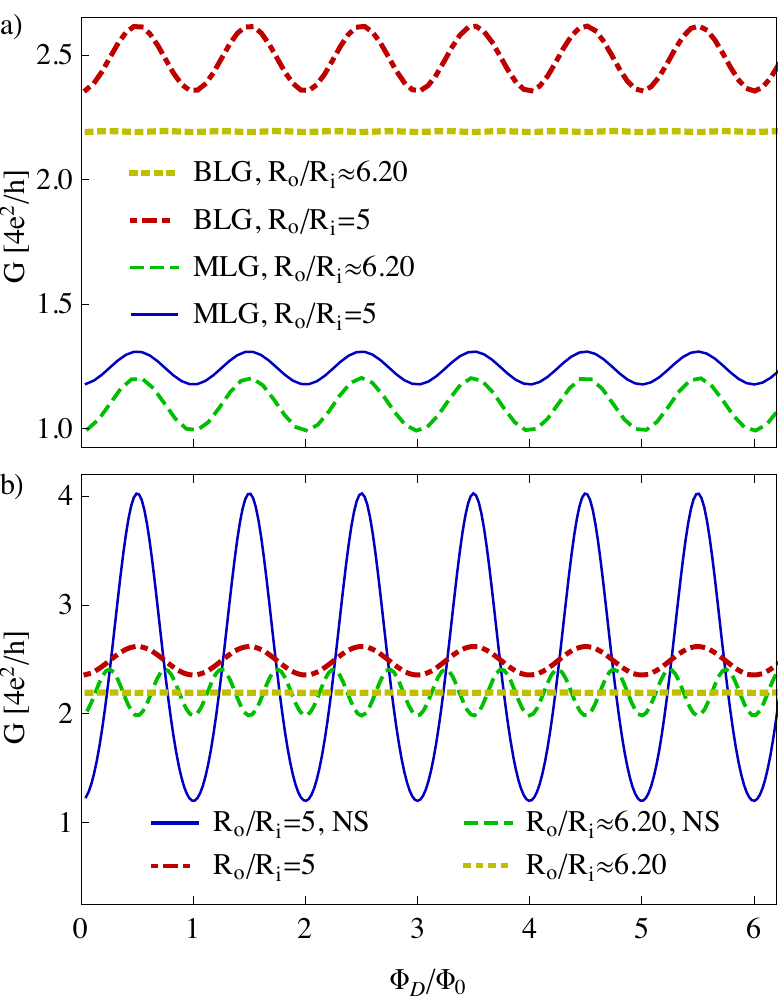}}
\caption{\label{tft3crop}
  Conductance of different graphene-based Corbino devices with the inner radius $R_{\rm i}=50\,l_{\perp}\simeq{}80\,$nm as a~function of the magnetic field. (a) Magnetoconductance oscillations in mono- and bilayer disks at the Dirac point for the two values of the radii ratio, for which the oscillation magnitude is close to the maximal ($R_{\rm o}/R_{\rm i}=5$) and to the minimal value ($R_{\rm o}/R_{\rm i}=6.2$) in the bilayer case. (b) Magnetoconductance of bilayer disks in normal Corbino and Andreev-Corbino (NS) setup. Notice the oscillation frequency doubling for the Andreev-Corbino setup and $R_{\rm o}/R_{\rm i}=6.2$. 
}
\end{figure}

\begin{figure}
\centerline{\includegraphics[width=0.9\linewidth]{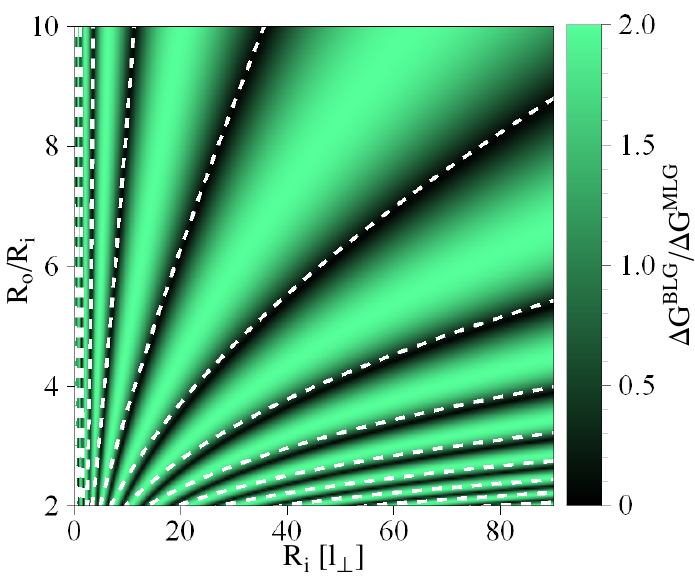}}
\caption{\label{corbmin}
  Oscillation magnitudes ratio for bilayer and monolayer disks, $\Delta{G}^{\rm BLG}/\Delta{G}^{\rm MLG}$ (where $\Delta{G}\equiv{}G_{\rm max}-G_{\rm min}$) . Dashed lines mark the parameter values obtained from the approximating Eq.\ (\ref{eq:condition}) for $p=1,3,5,\dots$.
}
\end{figure}

\subsection{Magnetoconductance at the Dirac point}
For $E=V=0$ and $|U_\infty|\rightarrow{}\infty$, transmission eigenvalues can be found analytically, and read
\begin{equation}
  \label{tmbipm}
  T_m^\pm = 
  \frac{1}{\cosh^2\left[\mathcal{L}(m\pm\mathcal{A}+\Phi_{D}/\Phi_{0})\right]},
\end{equation}
where $\Phi_{D}=\pi\left(R_{\rm o}^{2}-R_{\rm i}^{2}\right)B$ is the flux
piercing the disk area, $\mathcal{L}=\ln\left(R_{\rm o}/R_{\rm i}\right)$, and $\Phi_{0}=2\left(h/e\right)\mathcal{L}$. The parameter
\begin{equation}
  \mathcal{A}=-\frac{\mbox{ln}\left(\,\Upsilon-\sqrt{\Upsilon^{2}-1}\,\right)}{2\mathcal{L}},
\end{equation}
with $\Upsilon=\cosh(\mathcal{L})+\Lambda\sinh(\mathcal{L})$ and $\Lambda=(R_{\rm o}^{2}-R_{\rm i}^{2})/(4l_{\perp}^2)$, takes the values from the range of $1/2<{\cal A}<\infty$. Summing over the normal modes labeled by integer $m$, one immediately finds that $G$ (\ref{gland}) shows periodic oscillations as a~function of $\Phi_D$, with a~period equal to $\Phi_0$ (see Fig.\ \ref{tft3crop}), closely resembling the magnetoconductance behavior predicted for the Corbino disk in MLG \cite{Ryc10,Kat10}. However, for any fixed $\Phi_D$, Eq.\ (\ref{tmbipm}) describes the two transmission maxima  separated by a~distance of $2{\cal A}\,\hbar{}$ in the angular-momentum space. In turn, the corresponding contributions to the magnetoconductance may interfere constructively or destructively with each other. The nature of the interference depends both on the sample size and on the interlayer hopping integral $t_{\perp}$ \cite{tper0foo}. 

For a~clear overview of the effect, we represent $G$ following from Eqs.\ (\ref{gland}) and (\ref{tmbipm}) by a~Fourier series
\begin{equation}
  \label{eq:fourier}
  G=\frac{16g_{0}}{\mathcal{L}}+
  \sum_{q=1}^{\infty}G_{q}\cos\left(\frac{2\pi q\,\Phi_{D}}{\Phi_{0}}\right),
\end{equation}
where
\begin{multline}
  \label{fouramp}
  G_{q}=\frac{32\pi^2q\,g_{0}\cos\left(2\pi q\mathcal{A}\right)}{{\cal L}^2\sinh\left(\pi^{2}q/\mathcal{L}\right)} \\
  \equiv
  2(-)^qG_q^{\rm MLG}\cos\left(2\pi q\mathcal{A}\right), \ \ \ q=1,2,3,\dots .
\end{multline}
The constant term in Eq. (\ref{eq:fourier}), $16g_{0}/\mathcal{L}\equiv{}2G_{\it diff}^{\rm MLG}$ [see Eq.\ (\ref{gdiffmono})], gives the average conductance, which is simply twice as large as in the monolayer case \cite{Ryc10,Kat10}. Such a~sum-rule does not generically apply to the Fourier amplitudes $G_q$, which are related to the corresponding amplitudes for MLG ($G_q^{\rm MLG}$) via the second equality in Eq.\ (\ref{fouramp}). A~special case of $G_q=2G_q^{\rm MLG}$ occurs for ${\cal A}=1/2$ (see Ref.\ \cite{tper0foo}). For sufficiently large systems, we have $|G_1|\gg{}|G_2|\gg{}\dots$, and it is possible to find out approximate conditions for maximal and minimal oscillation magnitudes $\Delta{G}\equiv{}G_{\rm max}-G_{\rm min}\simeq{}2|G_1|$, namely
\begin{equation}
  \label{eq:condition}
  p\mathcal{L}\simeq{}4\ln\left(\frac{R_{\rm i}}{2l_{\perp}}\right),
\end{equation}
where $p$ is an even (odd) integers for maximal (minimal) oscillation magnitudes. As illustrated in Fig.\ \ref{corbmin}, the parameter values following from Eq.\ (\ref{eq:condition}) for odd $p$ [white dashed lines] coincides with the actual regions where the oscillation magnitude vanishes [black areas], provided that $R_{\rm o}/R_{\rm i}\gtrsim{}3$ and $R_{\rm i}/l_\perp\gtrsim{}10$. 

For the sake of completeness, we discuss now the magnetoconductance in the Andreev-Corbino setup, in which the disk-shaped sample is attached to one normal and one superconducting leads. In such a~situation, the conductance is given by \cite{Akh07}
\begin{equation}
  G^{\rm NS} = 2_s2_vg_{0}
 \mbox{Tr}\left[\,2\boldsymbol{T}^2(2-\boldsymbol{T})^{-2}\,\right].
\end{equation}
For BLG disk at the Dirac point, this leads to
\begin{multline}
  \label{gnssum}
  G^{\rm NS} = 
  8g_{0}\sum_{m=-\infty}^{\infty}\left\{\frac{1}{\cosh^{2}\left[2\mathcal{L}(
        \overline{m}+\mathcal{A})\right]} \right. \\ 
    \left. +\frac{1}{\cosh^{2}\left[ 2\mathcal{L}(
          \overline{m}-\mathcal{A}) \right]}\right\},
\end{multline}
where we have defined $\overline{m}=m+\Phi_D/\Phi_0$. $G^{\rm NS}$ (\ref{gnssum}) can be represented by a~Fourier series of the form given by Eq.\ (\ref{eq:fourier}) with the same average conductance ($2G_{\it diff}^{\rm MLG}$), and the amplitudes $G_{q}$ (\ref{fouramp}) replaced by 
\begin{equation}
  \label{fourans}
  G_q^{\rm NS}=\frac{16\pi^2q\,g_{0}\cos\left(2\pi q\mathcal{A}\right)}{{\cal L}^2\sinh\left[\pi^{2}q/(2\mathcal{L})\right]}.
\end{equation}
Strictly speaking, the scaling rule earlier found for the disk in MLG, namely $G_q^{\rm MLG, NS}({\cal L})=2G_q^{\rm MLG}(2{\cal L})$ \cite{Ryc10}, does not apply in the bilayer case due to the interlayer coupling manifesting itself via the ${\cal A}$-dependent factor in Eq.\ (\ref{fourans}). However, we still have $G_q^{\rm NS}/G_q\rightarrow{}1$ for $R_{\rm o}/R_{\rm i}\rightarrow{}\infty$ (and arbitrary $q$). Also, magnetoconductance oscillations for bilayer disks with moderate radii ratios  are noticeably amplified in the Andreev-Corbino setup in comparison to the normal Corbino setup [see Figs.\ \ref{tft3crop}(a) and \ref{tft3crop}(b)].

The approximate conditions for maximal and minimal oscillations, given by Eq.\ (\ref{eq:condition}), are essentially valid for both the normal Corbino and the Andreev-Corbino setup. The relation $|G_1^{\rm NS}|\gg|G_2^{\rm NS}|\gg|G_3^{\rm NS}|\gg\dots$ is satisfied, for a~moderate radii ratios, near the oscillations maxima [even $p$ in Eq.\ (\ref{eq:condition})], whereas close to the minima one typically gets $G_1^{\rm NS}\simeq{}0$ and $|G_2^{\rm NS}|\gg|G_3^{\rm NS}|\gg\dots$, leading to the visible oscillations frequency doubling [see Fig.\ \ref{tft3crop}(b)].  In the normal Corbino setup, with the radii fixed at $R_{\rm i}=50\,l_\perp$ and $R_{\rm o}=6.2\,R_{\rm i}$, the magnetoconductance  is almost constant [yellow dotted line]. On the other hand, if one of the leads is superconducting, the frequency of conductance oscillations is doubled in comparison to $\Phi_0^{-1}$ [green dashed line].

\begin{figure}
\centerline{\includegraphics[width=0.9\linewidth]{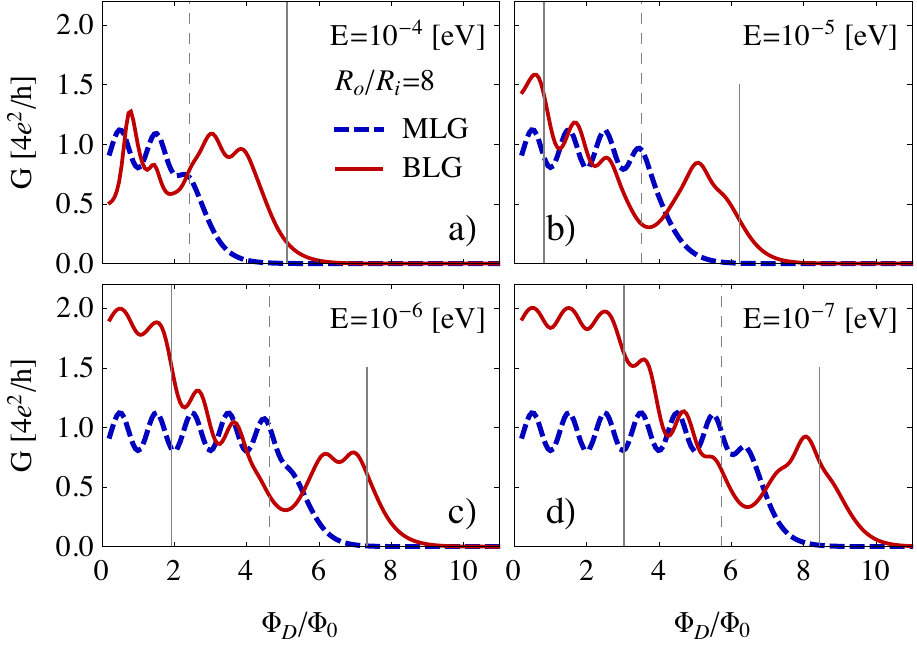}}
\caption{\label{grid2crop}
  Same as Fig.\ \ref{tft3crop}, but for the Fermi energy $E>0$ (specified at each panel). The radii ratio is fixed at $R_{\rm o}/R_{\rm i}=8$. Blue dashed and red solid lines correspond to the mono- and bilayer cases (respectively); the zero bias ($V=0$) is supposed for BLG. Vertical lines mark the values of $\Phi_D^{{\rm max},\,\eta}$ (\ref{phidmax}) for $\eta=0$ (grey dashed lines) and $\eta=\pm{}1$ (grey solid lines).
}
\end{figure}

\subsection{Finite-doping effects}
We now extend our analysis onto situations when the Fermi energy is close but not precisely adjusted to the Dirac point, keeping the zero bias between the layers ($E\neq{}0$, $V=0$). [Hereinafter, the normal-Corbino setup is considered.] The corresponding magnetoconductance spectra are presented in Fig.\ \ref{grid2crop}. In the monolayer case, the disk conductance at weak dopings follows the zero-doping curve for first few oscillation periods, and then starts to decrease rapidly with increasing field \cite{Ryc10} (see blue dashed lines in all panels of Fig.\ \ref{grid2crop}).  For BLG (see red solid lines) we have a~relatively wide crossover field interval, separating the oscillating and the field-suppressed conductance ranges. Typically, the conductance in the crossover interval does not decay monotonically with the field. Instead, a~well-defined magnetoconductance peak appears, with $G\simeq{}G_{\it diff}^{\rm MLG}$ near the maximum. Below, we link these features to the presence---in the vicinity of the Dirac point---of the two independent transmission channels for any angular momentum quantum number $m$, characterized by the transmission probabilities which are  numerically close to $T_m^\pm$ (\ref{tmbipm}).  

The contribution to the disk conductance originating from evanescent waves, for either MLG or BLG close to the Dirac point, can be roughly estimated by
\begin{equation}
  \sum_{l,\,{\sf eva}}T_l\sim{}
  \left(\frac{R_{\rm i}}{R_{\rm o}}\right)^{2|l_{\rm max}|}
  \ \ \ (\text{for }R_{\rm i}\ll{}R_{\rm o}),
\end{equation}
where $l_{\rm max}$ denotes the angular momentum corresponding to the maximal transmission at $E=0$, namely $l_{\rm max}=\eta{\cal A}-\Phi_D/\Phi_0$, where $\eta=0$ for MLG or $\eta=\pm{}1$ for BLG. The contribution from the propagating waves appearing for $E\neq{}0$ is of the order of
\begin{equation}
  \sum_{l,\,{\sf pro}}T_l\sim{}
  \left(k_0R_{\rm i}\right)^{2}\ \ \ (\text{for }k_0R_{\rm i}\ll{}1),
\end{equation}
where we have defined the wavevector $k_0=|E|/(\hbar{}v_F)$. 
Quasiperiodic magnetoconductance oscillations can be observed as long as $\sum_{l,\,{\sf eva}}T_l\gtrsim\sum_{l,\,{\sf pro}}T_l$, directly leading to the limits for magnetic fluxes
\begin{equation}
  \label{phidmax}
  |\Phi_D|\lesssim{}\Phi_D^{{\rm max},\,\eta}=\frac{2h}{e}
  \Big[ \,\eta{\cal A}{\cal L}-\ln\left(k_0R_{\rm i}\right)\, \Big].
\end{equation}
The values of $\Phi_D^{{\rm max},\,\eta}$, for $\eta=0,\pm{}1$, are also depicted in Fig.~\ref{grid2crop} (see vertical lines), showing that the flux range defined as $\Phi_D^{{\rm max},\,-1}\leqslant\Phi_D\leqslant\Phi_D^{{\rm max},\,+1}$ coincides with the crossover field interval for BLG disk with $R_{\rm i}=50\,l_\perp$, $R_{\rm o}/R_{\rm i}=8$, and $|E|\leqslant{}10^{-6}\,$eV. For larger $R_{\rm o}$, such a~coincidence can also be observed at higher $E$, provided that $\Phi_D^{{\rm max},\,-1}\gtrsim{}2\Phi_0$.

\begin{figure}
\centerline{\includegraphics[width=0.8\linewidth]{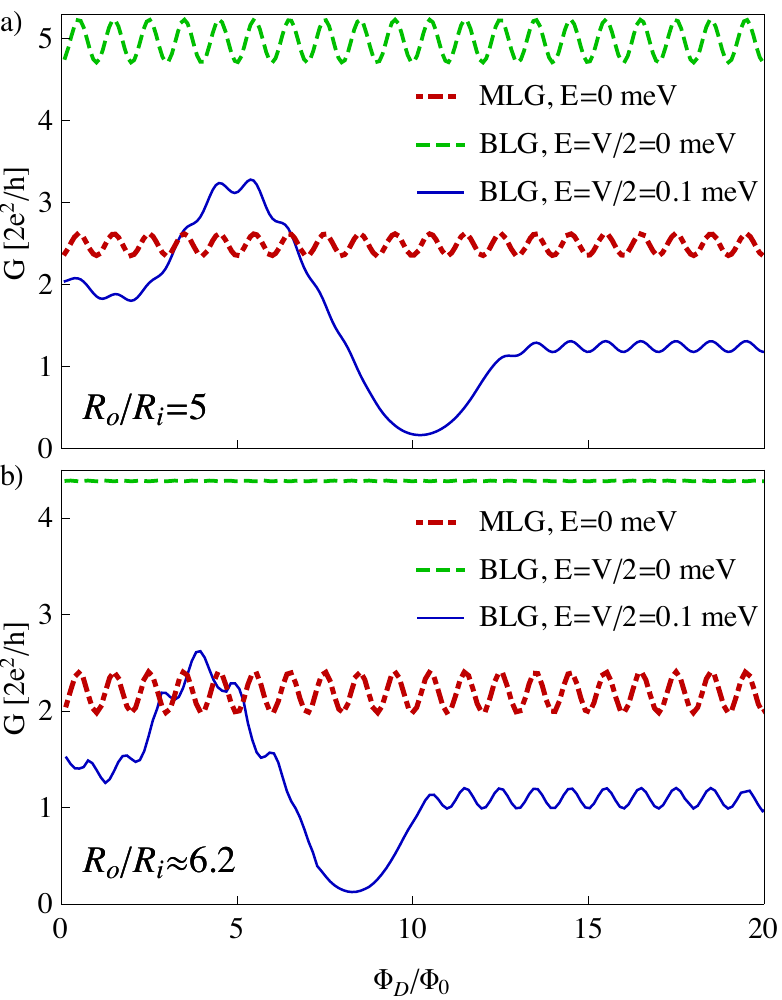}}
\caption{\label{delcrop}
  Same as Figs.\ \ref{tft3crop} and \ref{grid2crop}, but for the electrostatic bias between the layers $V/2=E=0.1\,$eV (blue solid lines). Remaining lines show the magnetoconductance spectra for the Corbino disk in unbiased and undoped BLG ($V/2=E=0$) [green dashed lines], as well as in undoped MLG [red dash-dot lines]. The values of the radii ratio are $R_{\rm o}/R_{\rm i}=5$ (top panel) and $R_{\rm o}/R_{\rm i}=6.2$ (bottom panel). 
}
\end{figure}

\subsection{The  biased sample case ($V\neq{}0$)}
We focus now on the effect of a~nonzero electrostatic bias between the layers in the normal-Corbino geometry. The corresponding magnetoconductance spectra for the two selected radii ratio  $R_{\rm o}/R_{\rm i}=5$ and  $R_{\rm o}/R_{\rm i}=6.2$ (with $R_{\rm i}=50\,l_\perp$) are presented in Fig.\ \ref{delcrop}, where we have fixed the Fermi energy at $E=V/2=0.1\,$eV. The disk conductance first shows rather irregular behavior with increasing field, varying in a~range of $0<G\lesssim{}G_{\it diff}^{\rm MLG}$ (the corresponding magnetoconductance spectra for $E=V=0$, and for undoped MLG disks, are also shown in Fig.\ \ref{delcrop}). For $\Phi_D\gtrsim{}10\,\Phi_0$, periodic oscillations are restored, but the average conductance is $4g_0/{\cal L}=G_{\it diff}^{\rm MLG}/2$. Also, the oscillations magnitude $\Delta{}G=\Delta{}G^{\rm MLG}$. (Notice that we have selected the disk radii such that $\Delta{}G^{\rm BLG}$ is close to the maximal and to the minimal value in the $E=V=0$ case, see green dashed lines.) These features can be attributed to the splittings of layer and valley degeneracies of the lowest Landau level in the presence of band gap and magnetic field (see Ref.\ \cite{Rut14a}).

Also for higher LLs, the disk conductance oscillates periodically with $\Phi_D$, qualitatively reproducing the behavior predicted for the monolayer case in Ref.\ \cite{Ryc10}. This is because finite doping eliminated the level degeneracy associated with the two layers, even in the absence of the electrostatic bias ($V=0$). For $V\neq{}0$, the valley degeneracy no longer applies, and the conductance further drops by a~factor of $2$. A complete overview of different transport regimes on the field-doping parameter plane is given in Sec.\ \ref{secon2deg}, where we compare (in a~quantitative manner) the magnetoconductance of the Corbino disks in BLG and in 2DEG.

\begin{figure}[!t]
\centerline{\includegraphics[width=\linewidth]{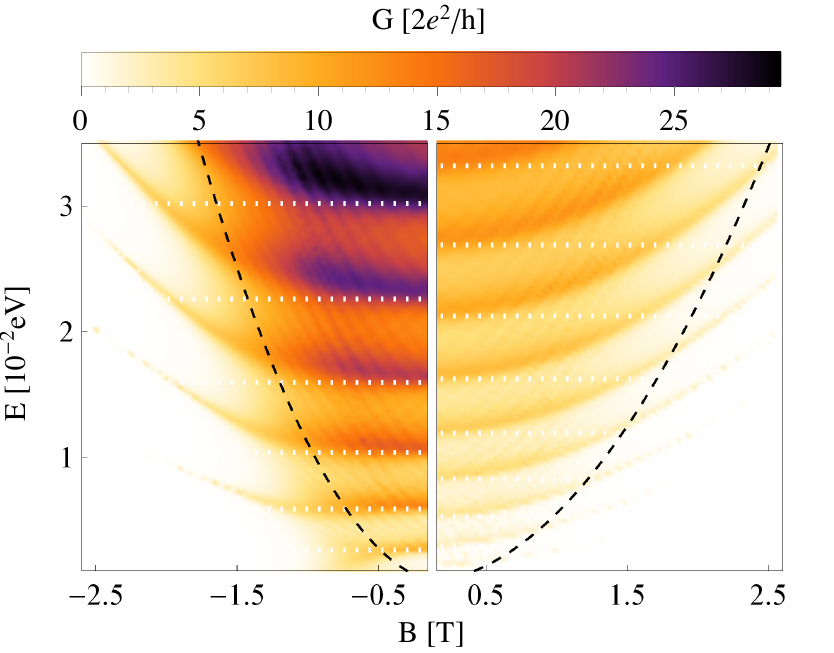}}
\caption{ \label{imash}
  Conductance as a~function of doping and magnetic field for the Corbino disks in unbiased BLG (left) and in 2DEG (right). The radii are fixed at $R_{\rm i}=25\,l_\perp\simeq{}40\,$nm and $R_{\rm o}=4R_{\rm i}$ for both cases. Black dashed lines mark the condition for cyclotronic diameters $2r_C=R_{\rm o}-R_{\rm i}$. White dotted lines depict the energy levels given by Eq.\ (\ref{qdotlevs}). 
}
\end{figure}

\begin{figure}
\centerline{\includegraphics[width=0.9\linewidth]{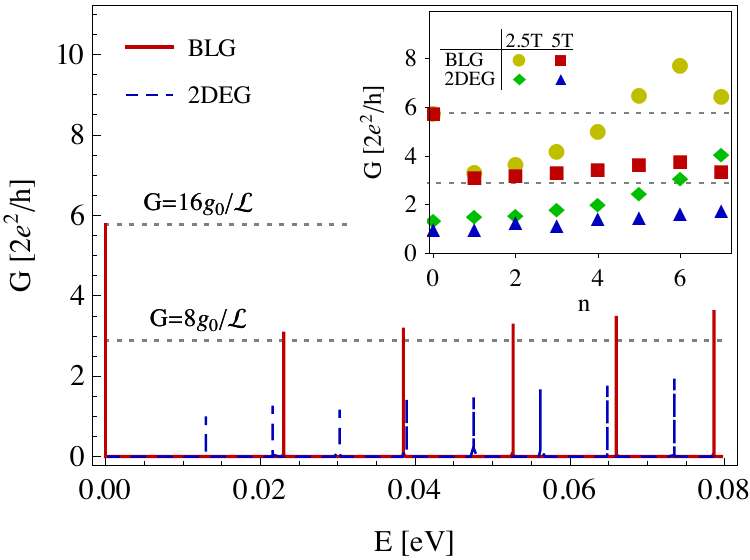}}
\caption{ \label{lp4v2}
  Conductance as a~function of doping at fixed $B=5\,$T. The parameters are same as in Fig.\ \ref{imash}. Inset shows the maximal conductance at the resonance with $n$-th LL, for the two different values of magnetic field ($B=2.5\,$T and $5\,$T). 
}
\end{figure}

\section{ \label{secon2deg}
  Magnetoconductance of the Corbino disk in 2DEG }

For both BLG and 2DEG systems, parabolic bands appear in the low-energy dispersion relation, and the effective masses are in the range of $m_{\star}/m_e=10^{-2}-10^{-1}$ (where $m_e$ denotes the free electron mass). Therefore, a~detailed comparison of the magnetic field effects described in Sec.\ \ref{seconqrce}, with analogues effects for the Corbino disk in 2DEG, is desired to identify the role of chiral tunneling of Dirac fermions in BLG. Below, we extend the mode-matching analysis presented in Ref.\ \cite{Ryc09} on the nonzero field situation. 

The effective Schr\"{o}dinger equation for electrons in 2DEG system reads
\begin{equation}
  \label{eq:ham-1}
\left[\frac{1}{2m_{\star}}\left(\frac{\hbar}{i}\mbox{\boldmath$\nabla$}+e{\bf A}\right)^{2}+U\left(r\right)\right]\psi=E\psi,
\end{equation}
where $\psi({\bf r})$ is the complex-scalar wavefunction, the vector potential ${\bf A}$ is same as in Eq.\ (\ref{eq:hamiltonian1}), and the Zeeman term is neglected again. The electrostatic potential energy $U(r)$ is still given by Eq.\ (\ref{eq:potential}), but we no longer assume infinite doping in the leads, as the mismatch in Fermi velocities results in zero transmission in such a~limit \cite{cor2degfoo,Kir94,Sou98,Pei79}. Instead, $U_\infty$ can be adjusted such that $\pi{}R_{\rm i}\sqrt{m_\star(E-U_\infty)/\hbar^2}\gtrsim{}10$, entering the multimode leads regime, in which the conductance only weakly depends on $U_\infty$. 

Since the Hamiltonian in Eq.\ (\ref{eq:ham-1}) commutes with the orbital momentum operator $L_z=-i\hbar\partial_{\phi}$, we choose wavefunctions of a form $\psi\left(r,\phi\right)=\varphi\left(r\right)\mbox{exp}\left(il\phi\right)$,
with $l$ integer. This bring us to solving the effective one-dimensional scattering problem, with the Sch\"{o}dinger equation
\begin{equation}
  \left[
    -\partial_{r}^{2}-\frac{1}{r}\partial_{r}
    +\frac{l^2}{r^2}+\frac{r^2}{4l_{B}^{4}}
  \right]\varphi\left(r\right)
  = \zeta_l\varphi\left(r\right),
\end{equation}
where $\zeta_l=2m_{\star}\left[E-U\left(r\right)\right]/\hbar^{2}-l/l_{B}^{2}$.
For the contact regions we have $l_B^{-1}=0$, and  the solutions are given by the Hankel functions \cite{Abr65a}, namely 
\begin{align}
  \varphi_{l}^{\rm (i)}(r) &= H_{l}^{(1)}(Kr)+r_lH_{l}^{(2)}(Kr), \nonumber\\
  \varphi_{l}^{\rm (o)}(r) &= t_lH_{l}^{(1)}(Kr), \label{phio2deg}
\end{align}
where $K=\sqrt{2m_{\star}(E-U_{\infty})/\hbar^{2}}$, $r_l$ ($t_l$) is the reflection (transmission) coefficient, and we have assumed scattering from the inner lead. 

For the disk area, we get
%
%
\begin{multline}
  \label{phd2deg}
  \varphi_l^{\rm (d)}(r)=
  \left(C_{l}/r\right)
  \mbox{W}_{\Omega_l,\,l/2}\left({\textstyle -\frac{1}{2}r^{2}/l_B^2}\right) \\
  +\left(D_{l}/r\right)
  \mbox{W}_{-\Omega_l,\,l/2}\left({\textstyle \frac{1}{2}r^{2}/l_B^2}\right), 
\end{multline}
where $\Omega_l=\left(l-k^2l_{B}^2\right)/2$ with $k=\sqrt{2m_\star{}E/\hbar^2}$, $W_{\kappa,\mu}(x)$ is the Whittaker function \cite{Abr65b},
and $C_l$, $D_l$ are arbitrary constants. In particular, imposing the normalization of $\varphi_l^{\rm (d)}$, one can obtain the well-known energy quantization
\begin{equation}
  \label{ll2deg}
  E_{n}=\hbar\omega_{c}\left(n+1/2\right),
\end{equation}
with $\omega_{c}=eB/\hbar$ and $n=0,1,2...$. For the open system studied here, normalization condition for wavefunctions do not apply, but LL energies $E_n$ (\ref{ll2deg}) coincides with the transmission maxima of $T_l=|t_l|^2$.

Carrying out the mode-matching procedure for each value of $l$ separately (see Appendix~C for the details), we get the Landauer-B\"{u}ttiker conductance $G=2_sg_0\sum_lT_l$ for arbitrary dopings and magnetic fields. For the numerical analysis, we set an effective mass same as in GaAs systems $m_{\star}=0.067m_{e}$, the inner radius is $R_{i}=25\,l_{\perp}\simeq{}40\,$nm and the doping on the leads is such that $E-U_{\infty}=0.4\,$eV. 

The results are displayed in Figs.\ \ref{imash} and \ref{lp4v2}. Both for BLG and 2DEG disks (see Fig.\ \ref{imash}) we observe, at low magnetic fields, well-defined conductance maxima corresponding to the quantum-dot energy levels
\begin{equation}
  \label{qdotlevs}
  E_q=
  \begin{cases}
    {\displaystyle\frac{1}{2}}
    \left[-t_\perp+\sqrt{t_\perp^2+\left(\frac{hv_F}{L}\right)^2q^2}\,\right]
    & \text{for BLG}, \\ 
    h^2q^2/\left(8m_\star{}L\right) & \text{for 2DEG}, \\ 
  \end{cases}
\end{equation}
with $L\equiv{}R_{\rm o}-R_{\rm i}$ and $q$ integer. These maxima gradually evolve, with increasing field, towards narrow peaks corresponding to the resonances with LLs, at energies given by Eq.\ (\ref{eq:llBLG}) for BLG or Eq.\ (\ref{ll2deg}) for 2DEG. Away from the maxima, some background conductance $G\gtrsim{}g_0$ appears when the cyclotronic diameter $2r_C\gtrsim{}L$. (Otherwise, $G\ll{}g_0$). In turn, the ballistic and the quantum-tunneling transport regimes can be identified for both the systems considered. 

The key difference in charge transport via Corbino disks in BLG and in 2DEG appears in the quantum-tunneling regime, and is visualized in Fig.\ \ref{lp4v2}. For BLG, the conductance at the local maximum corresponding to the resonance with $n$-th LL is $G_{\rm max}\simeq{}2G_{\it diff}^{\rm MLG}$ for $n=0$, or  $G_{\rm max}\simeq{}G_{\it diff}^{\rm MLG}$ for $n\neq{}0$. When increasing the magnetic field, every single resonance gets narrow in the energy scale, but the peak conductance is almost unaffected \cite{gpeakfoo}. To the contrary, transmission resonances for the disk in 2DEG simply vanishes with increasing field (see inset in Fig.\ \ref{lp4v2}), as the pseudodiffusive charge transport regime does not occur in this case.

\section{Conclusions}
We have investigated, by means of analytical mode-matching for the effective Dirac equation, the effects of the interlayer hopping and the electrostatic bias on magnetoconductance of the Corbino disk in bilayer graphene. Most remarkably, the disk conductance still shows periodic (approximately sinusoidal) oscillations  with the applied field, typically with the same period as in the monolayer case \cite{Ryc10,Kat10}, both when the system is at the Dirac point, or the values of electrochemical doping follows the field-dependent position of any higher Landau level at a~given bias. In any case, the average conductance coincides with the pseudodiffusive value for a~disk-shaped bilayer sample, provided the degeneracies associated with valley and layer degrees of freedom are correctly taken into account \cite{Rut14a}. A~quantitative comparison with a~similar system in 2DEG, for which the conductance gradually decays with increasing field, makes it clear that the chiral tunneling of Dirac fermions governs the charge transport through the Corbino disk in bilayer graphene. 

A special feature of the magnetoconductance spectra, directly linked to the presence of the hopping between the layers, may be observed for unbiased disk at the Dirac point. In such a~case, the two periodic contributions to the disk conductance may interfere constructively or destructively, depending on the geometric parameters (i.e., the radii $R_{\rm i}$, $R_{\rm o}$) and on the interlayer hopping integral ($t_\perp$). For particular combinations of these variables, obeying approximate Eq.\ (\ref{eq:condition}), which can be rewritten as
\begin{equation}
   \frac{R_{\rm o}}{R_{\rm i}}\simeq\left(
     \frac{R_{\rm i}\,t_{\perp}}{2\hbar{}v_{F}}
   \right)^{4/p}\ \ \ \text{with }p=1,3,5,\dots,
\end{equation}
the interference is maximally destructive, leading to the approximately field-independent conductance (twice as large as the pseudodiffusive value for the disk setup in a~monolayer \cite{Ryc09}) for moderate radii ratios $R_{\rm o}/R_{\rm i}\lesssim{}10$ in the normal Corbino setup, or to the oscillation frequency doubling for the Andreev-Corbino setup. We notice that the effect which we described offers, at lead in principle, an independent way of determining the basic microscopic parameters of bilayer graphene. 

Quite remarkably, the energy-gap opening by applying the external electrostatic bias affects transport properties of the Corbino disk in bilayer graphene in rather unexpected manner: New features, mentioned above and absent in a~monolayer case, appear for ubiased disks at the Dirac point, whereas the gap opening essentially reduces the variety of magnetotransport behaviors to the earlier described for monolayer disks. This observation seems particularly significant, as some experimental works showed that the energy gap may also appear spontaneously, due to electron-electron interactions, for bilayer samples close to the charge-neutrality point \cite{Rtt11,Bao12}. It must be noticed, however, that the results of other conductance measurements \cite{May11} coincide with theoretical predictions for unbiased bilayer, leaving an ambiguity concerning the role of interactions in the system. 

The effects of disorder \cite{Mac13}, lattice defects \cite{Dre10}, or magnetic impurities \cite{Cor09,Sun12}, which may modify transport properties of graphene-based devices, are beyond the scope of this paper as we have focussed on perfectly clean ballistic systems. Certain features of the results, including the fact that unit transmission appears periodically (for consequtive normal modes) with increasing field, and that the oscillation period is proportional to the ratio of fundamental constants $h/e$, allow us to believe that symmetry-protected quantum channels \cite{Wak07} would lead to magnetoconductance oscillations appearing in a~more general situation as well.

\section*{Acknowledgements}
We thank Marko Burkhard and Elham Moomivand for the correspondence. 
The work was supported by the National Science Centre of Poland (NCN) via Grant No.\ N--N202--031440, and partly by Foundation for Polish Science (FNP) under the program TEAM {\em ``Correlations and coherence in quantum materials and structures (CCQM)''}. G.R.\ acknowledges the support from WIKING project. Computations were partly performed using the PL-Grid infrastructure.


\appendix
\section{Wave functions}
In this Appendix we present the wavefunctions of charge carrier in bilayer graphene, having the form of eigenspinors of the total angular-momentum operator $J_z$ (\ref{jztotal}), and thus adjusted to study the scattering problem with a~cylindrical symmetry. The cases of zero and non-zero magnetic fields, relevant for the leads and the sample area in the system of Fig.\ \ref{bicorbino}, are discussed separately. 

\subsection{Zero magnetic field}
Four linearly-independent solutions of the Dirac equation $H\psi=E\psi$ with the Hamiltonian given by Eq.\ (\ref{eq:hamiltonian1}), corresponding to the angular-momentum quantum number $m$, have forms of envelope wavefunctions given by Eq.\ (\ref{eq:fi}). For $B=0$, radial parts of these functions can be written as
\begin{equation}
  \phi_{\sf in}^\pm(r)=
  \left(\begin{array}{c}
      H_{m}^{(1)}\!\left(s_{\pm}r\right) \\
      -s_{\pm}\,\epsilon_u^{-1} 
      H_{m-1}^{(1)}\!\left(s_{\pm}r\right) \\
      (\epsilon_u^{2}\!-\!\eta_{\pm})
      l_\perp\epsilon_u^{-1}
      H_{m}^{(1)}\!\left(s_{\pm}r\right)\\
      s_{\pm}(\epsilon_u^{2}\!-\!\eta_{\pm})
      l_\perp(\epsilon_u\epsilon_d)^{-1}
      H_{m+1}^{(1)}\!\left(s_{\pm}r\right)
\end{array}\right)
\end{equation}
for the waves propagating from $r=0$ (the index $\pm$ refers to the two subbands), or
\begin{equation}
  \phi_{\sf out}^\pm(r)=
  \left(\begin{array}{c}
      H_{m}^{(2)}\!\left(s_{\pm}r\right) \\
      -s_{\pm}\,\epsilon_u^{-1} 
      H_{m-1}^{(2)}\!\left(s_{\pm}r\right) \\
      (\epsilon_u^{2}\!-\!\eta_{\pm})
      l_\perp\epsilon_u^{-1}
      H_{m}^{(2)}\!\left(s_{\pm}r\right)\\
      s_{\pm}(\epsilon_u^{2}\!-\!\eta_{\pm})
      l_\perp(\epsilon_u\epsilon_d)^{-1}
      H_{m+1}^{(2)}\!\left(s_{\pm}r\right)
    \end{array}\right)
\end{equation}
for the waves propagating from $r=\infty$, with $s_\pm=\sqrt{\eta_\pm}$, $\eta_{\pm}=\left(\Delta^{2}+\tilde{\epsilon}^{2}\right)\pm\sqrt{\tilde{\epsilon}^{2}\left(4\Delta^{2}+1/l_{\perp}^{2}\right)-\Delta^{2}/l_{\perp}^{2}}$, $\epsilon_u=\tilde{\epsilon}+\Delta$, $\epsilon_d=\tilde{\epsilon}-\Delta$, $H_m^{(1)}(x)$ [$H_m^{(2)}(x)$] being the Hankel function of the first [second] kind \cite{Abr65a}, and the remaining symbols same as in Eq.\ (\ref{eq:diaceq}). For $\tilde{\epsilon}\rightarrow\infty$, the asymptotic forms of radial wavefunctions are \cite{asymhan}
\begin{equation}
  \label{phiinasm}
  \phi_{\sf in}^\pm(r)\simeq\sqrt{\frac{2}{\pi\tilde{\epsilon}\,r}}
  \,{\exp\!\left[{\textstyle
        i\left(\tilde{\epsilon}\,r\!-\!\frac{1}{2}\pi{}m\!-\!\frac{1}{4}\right)
      }\right]}
  \left(\begin{array}{c}
      1\\- i\\ \mp{}1\\ \pm{}i
    \end{array}\right) 
\end{equation}
and
\begin{equation}
  \label{phioutasm}
  \phi_{\sf out}^\pm(r)\simeq\sqrt{\frac{2}{\pi\tilde{\epsilon}\,r}}
  \,{\exp\!\left[{\textstyle
        -i\left(\tilde{\epsilon}\,r\!-\!\frac{1}{2}\pi{}m\!-\!\frac{1}{4}\right)
      }\right]}
  \left(\begin{array}{c}
      1\\ i\\ \mp{}1\\ \mp{}i
    \end{array}\right). 
\end{equation}

\subsection{Non-zero magnetic field}
At the Dirac point ($\varepsilon=\delta=0$) the radial part of the wavefunction, being a~general solution of Eq.\ (\ref{eq:diaceq1}), reads
\begin{multline}
  \label{phiddir}
  \phi_{\rm d}(r)=
  \alpha_{1}\left(\begin{array}{c}
      f_m(\rho)\\
      0\\
      0\\
      t\rho f_m(\rho)/2
    \end{array}\right)
  +\alpha_{2}\left(\begin{array}{c}
      0\\
      \rho^{-1}\bar{f}_m(\rho)\\
      0\\
      0
    \end{array}\right) 
  \\
  +\alpha_{3}\left(\begin{array}{c}
      0\\
      t\rho \bar{f}_m(\rho)/2\\
      \bar{f}_m(\rho)\\
      0
    \end{array}\right)
  +\alpha_{4}\left(\begin{array}{c}
      0\\
      0\\
      0\\
      \rho^{-1}f_m(\rho)
\end{array}\right),
\end{multline}
where $f_m(\rho)=\exp\left(-m\ln\!\rho-\rho^{2}/4\right)$, $\bar{f}_m(\rho)=1/f_m(\rho)$, $\alpha_{j}$ are arbitrary complex coefficients [taking different values depending whether the mode-matching analysis is carried out for the wave incoming from $r=0$ given by $\phi_{\sf in}^+(r)$ or  $\phi_{\sf in}^-(r)$; see Appendix~B], and the remaining symbols are same as in Eq.\ (\ref{eq:diaceq1}). 

At finite dopings ($\varepsilon\neq{}0$ or $\delta\neq{}0$), the radial wavefunctions are given by
\begin{equation}
  \label{philpm}
  \phi_l^\pm(r)=\left(\begin{array}{c}
      a_{m,l}^\pm(\varepsilon,\delta;\,\rho) \\
      b_{m,l}^\pm(\varepsilon,\delta;\,\rho) \\
      c_{m,l}^\pm(\varepsilon,\delta;\,\rho) \\
      d_{m,l}^\pm(\varepsilon,\delta;\,\rho)
    \end{array}\right)
  \ \ \ \ (\,l=1,2\,),
\end{equation}
where the spinor components can be written as
\begin{widetext}
\begin{align}
  \label{philpm1}
  a_{m,1}^{\pm}(\varepsilon,\delta;\,\rho) & =  
  2^{\left(m+1\right)/2}e^{\rho^{2}/4}\rho^{m}\,
\mbox{U}\left(\frac{\gamma_{\pm}}{2},m\!+\!1,-\frac{\rho^{2}}{2}\right), \nonumber \\
  b_{m,1}^{\pm}(\varepsilon,\delta;\,\rho) & =  
  -(\delta+\varepsilon)^{-1}\,2^{\left(m+1\right)/2}\,e^{\rho^{2}/4}\,\rho^{m-1}\,\left[\left(2m+\rho^{2}\right)\mbox{U}\left(\frac{\gamma_{\pm}}{2},m\!+\!1,-\frac{\rho^{2}}{2}\right) + \frac{\rho^{2}}{2}\gamma_{\pm}\mbox{U}\left(\frac{\gamma_{\pm}}{2}\!+\!1,m\!+\!2,-\frac{\rho^{2}}{2}\right)\right], \nonumber \\
  c_{m,1}^{\pm}(\varepsilon,\delta;\,\rho) & =  
 (\delta+\varepsilon)^{-1}\, 2^{\left(m+1\right)/2}e^{\rho^{2}/4}\,\rho^{m}\, t^{-1}\left[-\gamma_{\pm}+(\delta+\varepsilon)^{2}+2\right]\mbox{U}\left(\frac{\gamma_{\pm}}{2},m\!+\!1,-\frac{\rho^{2}}{2}\right), \nonumber \\
  d_{m,1}^{\pm}(\varepsilon,\delta;\,\rho) & =  
  (\delta^2-\varepsilon^2)^{-1}\,2^{\left(m-1\right)/2}e^{\rho^{2}/4}\,\rho^{m+1}\, t^{-1}\gamma_{\pm}\left[-\gamma_{\pm}+(\delta+\varepsilon)^{2}+2\right]
  \mbox{U}\left(\frac{\gamma_{\pm}}{2}\!+\!1,m\!+\!2,-\frac{\rho^{2}}{2}\right),
\end{align}
and
\begin{align}
  \label{philpm2}
  a_{m,2}^{\pm}(\varepsilon,\delta;\,\rho) & =  
  2^{\left(m+1\right)/2}e^{\rho^{2}/4}\rho^{m}\,\mbox{L}_{-\gamma_{\pm}/2}^{m}\left(-\frac{\rho^{2}}{2}\right),  \nonumber \\
  b_{m,2}^{\pm}(\varepsilon,\delta;\,\rho) & =  
  -(\delta+\varepsilon)^{-1}\,2^{\left(m+1\right)/2}\,e^{\rho^{2}/4}\,\rho^{m-1}\,\left[\rho^{2}\,\mbox{L}_{-\gamma_{\pm}/2-1}^{m+1}\left(-\frac{\rho^{2}}{2}\right) + \left(2m+\rho^{2}\right)\,\mbox{L}_{-\gamma_{\pm}/2}^{m}\left(-\frac{\rho^{2}}{2}\right)\right], \nonumber\\
  c_{m,2}^{\pm}(\varepsilon,\delta;\,\rho) & =  
  (\delta+\varepsilon)^{-1}\,2^{\left(m+1\right)/2}\,e^{\rho^{2}/4}\,\rho^{m}\,t^{-1}\left[-\gamma_{\pm}+(\delta+\varepsilon)^{2}+2\right] \frac{\Gamma\left(m-\gamma_{\pm}/2+1\right)}{\Gamma\left(1-\gamma_{\pm}/2\right)} \nonumber \\
  & \times {\cal F}\left(\frac{\gamma_{\pm}}{2}; m\!+\!1; -\frac{\rho^{2}}{2}\right), \nonumber \\
  d_{m,2}^{\pm}(\varepsilon,\delta;\,\rho) & =  
  (\delta^2-\varepsilon^2)^{-1}\,2^{\left(m+1\right)/2}\,e^{\rho^{2}/4}\,\rho^{m+1}\,t^{-1}\left[-\gamma_{\pm}+(\delta+\varepsilon)^{2}+2\right] \frac{\Gamma\left(m-\gamma_{\pm}/2+1\right)}{\Gamma\left(-\gamma_{\pm}/2\right)} \nonumber \\
  & \times{\cal F}\left(\frac{\gamma_{\pm}}{2}\!+\!1; m\!+\!2; -\frac{\rho^{2}}{2}\right).
\end{align}
We have further defined $\gamma_{\pm}=\left(\delta^{2}+\varepsilon^{2}\right)\pm\sqrt{\varepsilon^{2}\left(4\delta^{2}+t{}^{2}\right)-\delta^{2}t^{2}}$, $\mbox{L}_{b}^{a}\left(x\right)$ is the the generalized Laguerre polynomial \cite{Abr65c}, $\mbox{U}(a,b,x)$ denotes the confluent hypergeometric function \cite{Abr65b}, $\Gamma(z)=\int_0^\infty{}x^{z-1}e^{-x}dx$ is the Euler gamma function, ${\cal F}\left(a;b;z\right)\equiv{}_{1}\mbox{F}_{1}(a;b;z)\,\Gamma(b)$ with $_pF_q(a_1,\dots,a_p;b_1,\dots,b_q;z)$ denoting the generalized hypergeometric function \cite{Olv10}, and the remaining symbols are same as in Eq.\ (\ref{eq:diaceq1}) in the main text.

\section{Transmission eigenvalues}
The charge conservation conditions for the interfaces between the disk area and the leads ($r=R_{\rm i}$ and $r=R_{\rm o}$, see also Fig.\ \ref{bicorbino}) can be written, in terms of radial wavefunctions presented in Appendix~A, as 
\begin{equation}
  \label{eq:matching}
  \phi_{\sf in}^\pm(R_{\rm i}) 
  + r_p^\pm\phi_{\sf out}^+(R_{\rm i}) + r_n^\pm\phi_{\sf out}^-(R_{\rm i}) 
  = \phi_{\rm d}(R_{\rm i}), 
  \ \ \ \ 
  \phi_{\rm d}(R_{\rm o}) 
  = t_p^\pm\phi_{\sf in}^+(R_{\rm o}) + t_n^\pm\phi_{\sf in}^-(R_{\rm o}), 
\end{equation}
where we have represented wavefunctions in the leads following Eqs.\ (\ref{phiilin}) and (\ref{phiolin}) in the main text. In case the disk area is undoped and unbiased ($\varepsilon=\delta=0$), the function $\phi_{\rm d}(r)$ is given by Eq.\ (\ref{phiddir}). Taking the limit of $|U_\infty|\rightarrow\infty$ for the leads [i.e., choosing the functions  $\phi_{\sf in}^{\pm}(r)$, $\phi_{\sf out}^{\pm}(r)$ as given by Eqs.\ (\ref{phiinasm}) and (\ref{phioutasm})] and solving the system of linear equations following from Eq.\ (\ref{eq:matching}), one gets the closed-form expression for $T_m^\pm$ transmission eigenvalues for a~given angular momentum quantum number $m$  [see Eq.\ (\ref{tmbipm}) in the main text]. 

For a~more general case of finite dopings in the disk area ($\varepsilon\neq{}0$ or $\delta\neq{}0$) the limit of  $|U_\infty|\rightarrow\infty$ for the leads, combined with radial wavefunctions of the form $\phi_1^\pm(r)$,  $\phi_2^\pm(r)$ [see Eqs.\ (\ref{philpm}), (\ref{philpm1}), and (\ref{philpm2})] for the disk area, bring us to the system of linear equations 
\begin{equation}
  \label{eq:system8}
    \left(\begin{array}{cccccccc}
          -1 & -1 & a_{m,1}^{+}(\varepsilon,\delta;\,\rho_0) & a_{m,2}^{+}(\varepsilon,\delta;\,\rho_0) & a_{m,1}^{-}(\varepsilon,\delta;\,\rho_0) & a_{m,2}^{-}(\varepsilon,\delta;\,\rho_0) & 0 & 0\\
        -i & -i & b_{m,1}^{+}(\varepsilon,\delta;\,\rho_0) & b_{m,2}^{+}(\varepsilon,\delta;\,\rho_0) & b_{m,1}^{-}(\varepsilon,\delta;\,\rho_0) & b_{m,2}^{-}(\varepsilon,\delta;\,\rho_0) & 0 & 0\\
        1 & -1 & c_{m,1}^{+}(\varepsilon,\delta;\,\rho_0) & c_{m,2}^{+}(\varepsilon,\delta;\,\rho_0) & c_{m,1}^{-}(\varepsilon,\delta;\,\rho_0) & c_{m,2}^{-}(\varepsilon,\delta;\,\rho_0) & 0 & 0\\
        i & -i & d_{m,1}^{+}(\varepsilon,\delta;\,\rho_0) & d_{m,2}^{+}(\varepsilon,\delta;\,\rho_0) & d_{m,1}^{-}(\varepsilon,\delta;\,\rho_0) & d_{m,2}^{-}(\varepsilon,\delta;\,\rho_0) & 0 & 0\\
        0 & 0 & a_{m,1}^{+}(\varepsilon,\delta;\,\rho_1) & a_{m,2}^{+}(\varepsilon,\delta;\,\rho_1) & a_{m,1}^{-}(\varepsilon,\delta;\,\rho_1) & a_{m,2}^{-}(\varepsilon,\delta;\,\rho_1) & -{\cal R} & -{\cal R} \\
        0 & 0 & b_{m,1}^{+}(\varepsilon,\delta;\,\rho_1) & b_{m,2}^{+}(\varepsilon,\delta;\,\rho_1) & b_{m,1}^{-}(\varepsilon,\delta;\,\rho_1) & b_{m,2}^{-}(\varepsilon,\delta;\,\rho_1) & i{\cal R} & i{\cal R} \\
        0 & 0 & c_{m,1}^{+}(\varepsilon,\delta;\,\rho_1) & c_{m,2}^{+}(\varepsilon,\delta;\,\rho_1) & c_{m,1}^{-}(\varepsilon,\delta;\,\rho_1) & c_{m,2}^{-}(\varepsilon,\delta;\,\rho_1) & {\cal R} & -{\cal R} \\
        0 & 0 & d_{m,1}^{+}(\varepsilon,\delta;\,\rho_1) & d_{m,2}^{+}(\varepsilon,\delta;\,\rho_1) & d_{m,1}^{-}(\varepsilon,\delta;\,\rho_1) & d_{m,2}^{-}(\varepsilon,\delta;\,\rho_1) & -i{\cal R} & i{\cal R}
      \end{array}\right)
    \left(\begin{array}{c}
        r_p^{\pm}\\
        r_n^{\pm}\\
        \alpha_1^{\pm}\\
        \alpha_2^{\pm}\\
        \alpha_3^{\pm}\\
        \alpha_4^{\pm}\\
        t_p^{\pm}\\
        t_n^{\pm}
      \end{array}\right)=
    \left(\begin{array}{c}
         1 \\
        - i \\
        \mp1 \\
       \pm i \\
        0\\
        0\\
        0\\
        0
      \end{array}\right),
\end{equation}
with $\rho_0=R_{\rm i}/l_B$, $\rho_1=R_{\rm o}/l_B$, and ${\cal R}=\sqrt{R_{\rm i}/R_{\rm o}}$. The elements of reflection and transmission matrices ${\bf\tilde{r}}_{K,m}$,  ${\bf\tilde{t}}_{K,m}$ occurring in Eq.\ (\ref{eq:system8}) differ from the corresponding elements of ${\bf r}_{K,m}$,  ${\bf t}_{K,m}$ [see also Eq.\ (\ref{rt1block}) in the main text] only via phase factors, which are insignificant when calculating transmission eigenvalues. Solving Eq.\ (\ref{eq:system8}), one obtains the matrices ${\bf\tilde{r}}_{K,m}$, and ${\bf\tilde{\bf t}}_{K,m}$ for the $K$ valley and the angular momentum quantum number $m$. The reflection and transmission matrices for the $K'$ valley can be obtained from an~analogous procedure, starting from radial wavefunctions modified according to $(\phi_1,\phi_2,\phi_3,\phi_4)^T\rightarrow(\phi_1,-\phi_2,\phi_3,-\phi_4)^T$, with a~substitution $\delta\rightarrow{}-\delta$.

\section{Mode-matching for the disk in 2DEG}
For Schr\"{o}dinger electrons in the Corbino setup, the current conservation at $r=R_{\rm i}$ and  $r=R_{\rm o}$ leads to four independent matching conditions
$$
\varphi_l^{\rm (i)}(R_{\rm i})=\varphi_l^{\rm (d)}(R_{\rm i}), \ \ \ \
\varphi_l^{\rm (d)}(R_{\rm o})=\varphi_l^{\rm (o)}(R_{\rm o}),
$$
\begin{equation}
  \left.\frac{d\varphi_l^{\rm (i)}}{dr}\right|_{R_{\rm i}}=
  \left.\frac{d\varphi_l^{\rm (d)}}{dr}\right|_{R_{\rm i}}, \ \ \ \ 
  \left.\frac{d\varphi_l^{\rm (d)}}{dr}\right|_{R_{\rm o}}=
  \left.\frac{d\varphi_l^{\rm (o)}}{dr}\right|_{R_{\rm o}},
\end{equation}
determining the coefficients $r_l$, $t_l$, $C_l$, and $D_l$, defined via Eqs.\ (\ref{phio2deg}) and (\ref{phd2deg}) in the main text. Let us further define the wavefunctions in the disk area
\begin{equation}
  \varphi_{1,l}^{(\rm d)}\left(r\right)=
  \frac{1}{r}\,\mbox{W}_{\Omega_l,l/2}\left(-\frac{r^{2}}{2l_{B}^2}\right),
  \ \ \ \text{and} \ \ \ 
  \varphi_{2,l}^{(\rm d)}\left(r\right)=
  \frac{1}{r}\,\mbox{W}_{-\Omega_l,l/2}\left(\frac{r^{2}}{2l_{B}^2}\right),
\end{equation}
where $\Omega_l=\left(l-k^2l_{B}^2\right)/2$, $k=\sqrt{2m_\star{}E/\hbar^2}$, and $W_{\kappa,\mu}(x)$ is the Whittaker function \cite{Abr65b}. [\,In turn, $\varphi_l^{(\rm d)}(r)\equiv{}C_l\varphi_{1,l}^{(\rm d)}\left(r\right)+D_l\varphi_{1,l}^{(\rm d)}\left(r\right)$.\,] The transmission probability for the angular momentum quantum number $l$ can now be written as
\begin{equation}
  T_l=|t_l|^2=\frac{1}{|{\cal M}_l|^2}
  \left(\frac{4}{\pi{}l_B^2{}R_{\rm i}R_{\rm o}}\right)^2,
\end{equation}
where 
\begin{multline}
\mathcal{M}_l  =  \mbox{H}_{l}^{\left(1\right)}\left(KR_{\rm i}\right)\mbox{H}_{l}^{\left(2\right)}\left(KR_{\rm o}\right)\left[\partial_{r}\varphi_{1,l}^{({\rm d})}\left(R_{\rm o}\right)\partial_{r}\varphi_{2,l}^{({\rm d})}\left(R_{\rm i}\right)-\partial_{r}\varphi_{1,l}^{({\rm d})}\left(R_{\rm i}\right)\partial_{r}\varphi_{2,l}^{({\rm d})}\left(R_{\rm o}\right)\right] \\
    +K^2\left[\partial_{\rho}\mbox{H}_{l}^{\left(1\right)}\left(KR_{\rm i}\right)\right]\left[\partial_{\rho}\mbox{H}_{l}^{\left(2\right)}\left(KR_{\rm o}\right)\right]\left[\varphi_{1,l}^{({\rm d})}\left(R_{\rm o}\right)\varphi_{2,l}^{({\rm d})}\left(R_{\rm i}\right)-\varphi_{1,l}^{({\rm d})}\left(R_{\rm i}\right)\varphi_{2,l}^{({\rm d})}\left(R_{\rm o}\right)\right] \\
  +K\mbox{H}_{l}^{\left(1\right)}\left(KR_{\rm i}\right)\left[\partial_{\rho}\mbox{H}_{l}^{\left(2\right)}\left(KR_{\rm o}\right)\right]\left[\partial_{r}\varphi_{1,l}^{({\rm d})}\left(R_{\rm i}\right)\varphi_{2,l}^{({\rm d})}\left(R_{\rm o}\right)-\varphi_{1,l}^{({\rm d})}\left(R_{\rm o}\right)\partial_{r}\varphi_{2,l}^{({\rm d})}\left(R_{\rm i}\right)\right] \\
    +K\left[\partial_{\rho}\mbox{H}_{l}^{\left(1\right)}\left(KR_{\rm i}\right)\right]\mbox{H}_{l}^{\left(2\right)}\left(KR_{\rm o}\right)\left[\varphi_{1,l}^{({\rm d})}\left(R_{\rm i}\right)\partial_{r}\varphi_{2,l}^{({\rm d})}\left(R_{\rm o}\right)-\partial_{r}\varphi_{1,l}^{({\rm d})}\left(R_{\rm o}\right)\varphi_{2,l}^{({\rm d})}\left(R_{\rm i}\right)\right],
\end{multline}
and the derivatives are given by
\begin{align}
  \partial_{\rho}\mbox{H}_{l}^{\left(\alpha\right)}\left(\rho\right) & =  
  \mbox{H}_{l-1}^{\left(\alpha\right)}\left(\rho\right)
  -\frac{l}{\rho}\,\mbox{H}_{l}^{\left(\alpha\right)}\left(\rho\right), \\
  \partial_{r}\varphi_{\alpha,l}^{({\rm d})}\left(r\right) & =  
  -\frac{1}{r^{2}}\left[ \left(2\lambda_\alpha\Omega_l+1+\frac{\lambda_\alpha{}r^2}{2l_B^2}\right)\mbox{W}_{\lambda_\alpha\Omega_l,\,l/2}\left(-\frac{\lambda_\alpha r^{2}}{2l_{B}^2}\right) 
    + 2\mbox{W}_{1+\lambda_\alpha\Omega_l,\,l/2}\left(-\frac{\lambda_\alpha r^{2}}{2l_{B}^2}\right)\right] ,
\end{align}
for $\alpha=1,2$, and $\lambda_\alpha=-(-1)^\alpha$.

\end{widetext}


\begin{thebibliography}{}
\bibitem{footriwa}
Skew-interlayer hoppings may lead to the appearance of secondary Dirac points affecting transport properties of unbiased samples, but a gap still opens when applying the electrostatic field, see Ref.\ \cite{Mac13}.
\bibitem{tper0foo}
In the limit of $t_\perp\rightarrow{}0$, corresponding the separation of two layers, we have $l_\perp\rightarrow\infty{}$ and ${\cal A}\rightarrow{}1/2$, leading to the same conductance per layer as in the monolayer case.
\bibitem{cor2degfoo}
We further notice that the early theoretical works on the Corbino disks in 2DEG \cite{Kir94,Sou98} utilized a~particular form of the effective radial potential, leading to the conductance quantization. The quantization steps are smeared out when employing the direct mode-matching technique (see Ref.\ \cite{Ryc09}), leading to yet another surprise when discussing the applicability of the adiabatic expansion to scattering problems in physics \cite{Pei79}.
\end{thebibliography}
\end{document}